\documentclass[final, 3p, times, twocolumn, authoryear]{elsarticle}

\usepackage{amssymb}
\usepackage{amsthm}

\usepackage{natbib}
\usepackage[utf8]{inputenc} 
\usepackage[T1]{fontenc}    
\usepackage{url}            
\usepackage{booktabs}       
\usepackage{amsfonts}       
\usepackage{nicefrac}       
\usepackage{microtype}      
\usepackage[colorlinks]{hyperref}

\usepackage{subcaption}
\usepackage{enumitem}

\usepackage{lineno} 

\usepackage[misc,geometry]{ifsym}

\usepackage{verbatim}
\usepackage{array}
\usepackage{color}
\usepackage{caption}
\usepackage{multirow}
\usepackage[table]{xcolor}

\usepackage{csquotes}

\usepackage{pdflscape}

\journal{IFAC2023, Yokohama, Japan}

\renewcommand{\figureautorefname}{Figure~}
\renewcommand{\sectionautorefname}{Section~}

\begin{document}
\begin{frontmatter}

\title{Implementation of Autonomous Supply Chains for Digital Twinning: a Multi-Agent Approach\tnoteref{titlemark}}
\tnotetext[titlemark]{This paper has been accepted by the IFAC World Congress 2023, Yokohama, Japan, and will be published in IFAC-PapersOnLine: \url{https://www.sciencedirect.com/journal/ifac-papersonline}}

\author[first]{Liming Xu}
\ead{lx249@cam.ac.uk}
\author[first]{Yaniv Proselkov}
\ead{yp289@cam.ac.uk}
\author[first]{Stefan Schoepf}
\ead{ss2823@cam.ac.uk}
\author[second]{David Minarsch\fnref{david}}
\ead{david.minarsch@fetch.ai}
\fntext[david]{Now at Valory AG, Zug, Switzerland.}
\author[second]{Maria Minaricova}
\ead{maria.minaricova@fetch.ai}
\author[first]{Alexandra Brintrup}
\ead{ab702@cam.ac.uk}

\address[first]{Supply Chain AI Lab, Institute for Manufacturing, Department of Engineering, University of Cambridge, Cambridge, CB3 0FS, UK}

\address[second]{Fetch.ai, St John's Innovation Centre, Cowley Rd, Cambridge, CB4 0WS, UK}

\begin{abstract} 
Trade disruptions, the pandemic, and the Ukraine war over the past years have adversely affected global supply chains, revealing their vulnerability.
Autonomous supply chains are an emerging topic that has gained attention in industry and academia as a means of increasing their monitoring and robustness. While many theoretical frameworks exist, there is only sparse work to facilitate generalisable technical implementation.
We address this gap by investigating multi-agent system approaches for implementing autonomous supply chains, presenting an autonomous economic agent-based technical framework. 
We illustrate this framework with a prototype, studied in a perishable food supply chain scenario, and discuss possible extensions.  
\end{abstract}


\end{frontmatter}
\section{Introduction}\label{sec:introduction}
Recent trade disruptions, the pandemic, and the Ukraine war over the past years have seriously revealed  vulnerabilities of traditional global supply chains (SCs) \citep{handfield2020corona, shih2020global, kilpatrick2022supply} reinforcing the need for organisations to establish more resilient SCs.

The use of digital twin technologies has been widely discussed as a solution to manage disruptions and achieve more resilient SCs \citep{calatayud2019self, blueyonder2020autonomous, nelsonhall2021moving}.
Digital Twins (DT)  are virtual representations of physical objects and processes that allow real-time analytics and corrective actions \citep{sharma2022digital}. 
As such, DTs can be viewed as self-interested agents acting in pursuit of their goals. 
The manufacturing industry embraced DT solutions in areas such as process control and condition-monitoring, however, end-to-end Supply Chain DT remain a less well-established area. Whilst DT allows manufacturers to achieve better outcomes and  automate operational decisions at scale, a crucial element remains unaddressed to extend the concept to a SC context: A supply chain is a business entity group involved in the upstream and downstream flows of materials, information, and finance from sources to customers \citep{christopher2016logistics}. 
Thus, supply chains consist of multiple, heterogeneous agents that are interdependent. 

When the DT of a particular company adjusts batch sizes to minimise inventory costs, this action would impact downstream logistics processes by having to schedule additional deliveries, increasing carbon footprint and delivery costs. In such a scenario, system-level goals need trade-offs between individual agent goals, yet no reusable frameworks exist to facilitate this.
We postulate that the realistic applicability of the DT concept to SC will crucially depend on their ability to function in a multi-agent environment, capable of distributed decision-making for attaining system-level goals. 

A relevant development is the multi-agent system-based facilitation of Autonomous Supply Chains (ASC),  formally conceptualised in \citet{xu2022autonomous}. 
\citet{xu2022autonomous} has proposed that the natural extension of a DT operating in a supply chain would be to give it agency so that consensus among distributed stakeholders can be built. Equipping DT with a multi-agent system framework would integrate different supply chain functions and actors' data coherently and enable collective decision-making. While \citet{xu2022autonomous} presented the theoretical background of the ASC, the technical framework for implementing an ASC has not been well studied.

This paper addresses this gap, exploring the development of a technical framework for realising ASC systems for supply chain digital twinning. Generalisability is particularly important as current academic research on supply chain automation focus on siloed SC functions, such as pallet picking and demand forecasting and remains disconnected from studies in Digital Twinning.  

The main contributions of this paper are described below:
\begin{enumerate}[nosep]
    \item We demonstrate the use of multi-agent systems (MAS) in supply chains to facilitate an ASC for supply chain digital twinning.
    \item We extend the open economic framework (OEF) \citep{minarsch2021autonomous} and autonomous economic agent-based technical framework for agent-based autonomous supply chain implementation.
    \item We develop an A2SC prototype to demonstrate the MAS approach for building a prototype ASC.
\end{enumerate}
This paper is organised as follows:
\sectionautorefname \ref{sec:related_work} reviews work on using the MAS approach for supply chain management (SCM). 
\sectionautorefname \ref{sec:a2sc} presents the agent-based autonomous supply chain. 
An implementation of an A2SC prototype is presented in \sectionautorefname\ref{sec:prototype}. 
\sectionautorefname\ref{sec:discussion} discusses the limitations and implications of this work and concludes this paper.

\section{Related Work}\label{sec:related_work}
The use of MAS approaches in SCM originates in the Integrated SCM System (ISCM) developed by \citet{fox1993integrated} in the early 1990s. 
The ISCM was composed of a set of software agents, each responsible for one or more activities in the SC and coordinated with other agents to plan and execute SC functions.
\citet{fox2001agent} investigated key issues and presented solutions for constructing such an agent-based SC architecture in a detailed manner. 
\citet{swaminathan1998modeling} modelled SC dynamics with a MAS approach. 
They developed a software agent library that captured generic SC processes and concepts, providing a modular, reusable framework for developing a wide range of realistic SC models. 
In comparison with \citet{fox1993integrated, fox2001agent} which decomposed agents along the dimension of SC functions, \citet{swaminathan1998modeling} defined agents by considering the SC structural elements; the proposed framework in \citet{swaminathan1998modeling} allows a model to address issues at both tactical (coordination) and strategic levels (configuration). 
These early works paved the way to tackle SCM issues with MAS approaches.

Since these early studies, numerous literature on MAS-based SCM has emerged. 
Work on the use of MAS approaches for automating SCM can be classified into three main streams. 
The first stream tackles agent frameworks and architectures for generic or specific SCs \citep{sadeh2001mascot, kumar2013multi}.
The second stream tackles dynamic SC formation and configuration \citep{kim2010supply, ameri2013multi}. 
The third stream tackles the decision-making aspects of coordination and negotiation, making individual agents work coherently in a SCM \citep{sadeh2001mascot, wong2010multi}.
Other peripheral work has enhanced specific SC functions via agent technology, such as  
demand forecasting \citep{liang2006agent} and
supplier selection \citep{ghadimi2018multi}.
\citet{xu2021will} contains a more comprehensive review of MAS for SCM. 
These previous works showed the wide recognition of the suitability of MAS approaches for SCM, work has mostly focussed on employing MAS approaches to solve particular SCM issues rather than developing generalisable architectures.

While the MAS paradigm in SCM mostly focuses on either simulation or real-time control of supply chain operations, the concept of DTs in SCM is still gaining traction, with various definitions emerging. 
Most applications are in the area of logistics, and warehouse monitoring, although some authors classify all manufacturing processes, shop-floor, or even post-sales condition monitoring DT under the SC banner \citep{nguyen2022knowledge}.  
\citet{rosen2015importance} highlights that while Digital Twins should in theory be closely linked to autonomous systems research, the link has not been much explored. 
In SCM applications, the links to autonomous decision-making and actuation are at times implicitly made \citep{rosen2015importance, sharma2022digital, lopez2011resource}, where the argument is that as autonomous systems need real-time information, and an up-to-date DT would underline the autonomous system's function. 
However, in the majority of SC cases, the link to autonomy is not discussed at all \citep{nguyen2022knowledge}, with the DT only scoped as a monitoring and decision-aid tool. 
In this paper, we take the view that the DT is an ideal framework for feeding real-time information on the system state to a set of agents operating the supply chain. 
Similarly, a DT will enable monitoring of resulting states through the actions taken by agents, creating a continuous feedback loop.  
Thus the two concepts --- agent and DT --- are closely interlinked. 
In this paper, we implemented a scenario, in which we deploy a DT via an IoT system to track the system state, which is fed them into an autonomous SC agent framework.

\section{Agent-based Autonomous Supply Chain}\label{sec:a2sc}
Here we describe and categorise four key issues that arise when a multi-agent system (MAS) approach is applied to a supply chain system. A MAS is a {\it loosely coupled} network of software agents that interact to solve complex problems beyond each agent's individual capacities or knowledge. 
Each agent is a goal-directed computational entity that acts autonomously on behalf of its users, communicating and coordinating with other agents as needed.
As presented in \citet{xu2022autonomous}, an ASC is a set of self-interested organisational entities with partial knowledge that autonomously manage the movement of material and information flows.
A MAS framework is thus naturally suited for studying SCs \citep{fox1993integrated,  swaminathan1998modeling}. 
We explore the use of the MAS approach for building up ASCs, connecting distributed automated functions/entities with each other, and making them work coherently.
The resulting ASCs are called agent-based autonomous supply chains, or A2SCs.

Four design issues must be addressed when constructing A2SC. \citep{smith1981frameworks, fox2001agent}. 
The \emph{first} is the decomposition and distribution of task processing responsibility across agents. 
This requires the problem under study to be  partitioned and allocated to constituent agents that are logically and often geographically distributed. 
Appropriate decomposition can facilitate interaction among these agents, addressing the {\it second} issue.
The agents in a MAS resolve conflicts (e.g., via negotiation) or act coherently (e.g., via coordination) through interaction. 
In the SCM domain, both control and data are distributed hence there is neither global control nor global data storage.
Loosely coupled agents must connect with appropriate ones, proactively or reactively, to obtain control of or access to data.
A well-designed connection mechanism lowers communication overhead, reducing the amount of data transmitted and promoting responsiveness.

The {\it third} issue is the design of appropriate communication languages and interaction protocols, which is an essential enabler for collective decision-making.
Distributed agents require common languages to communicate with each other, mutually exchange knowledge, and need protocols to regulate their interactions. 
Standard language and protocols provide agents with instruments to establish connections and share information.
Representative work (still dominant and in current use) on these two aspects include 
FIPA-ACL \citep{fipa2002agent}, 
and the contract-net protocol \citep{smith1981frameworks}.

The {\it fourth} issue is vocabulary and knowledge representation.
To understand the contents of a message agents must use {\it consistent} words that refer to the objects, functions, and relations appropriate to a certain application they are familiar with. Thus, in addition to ``consistent words'', a commonly understood knowledge representation is an integral part of agent communication. 
A vocabulary may contain multiple {\it ontologies} for describing the shared body of knowledge in a domain.

The first two issues deal with conceptual aspects: the {\it static structure} and the {\it dynamics} of a MAS as they are relevant to the architecture and agent organisation of the resulting MAS, whereas the last two issues tackle more concrete aspects --- the tools to enable the connection between disparate agents. 
Communication languages and protocols allow agents to freely join in or exit from an operating MAS, enabling an open, dynamic and adaptive computational environment. 
These four issues presented above thus offer a guiding technical framework for designing an A2SC system. 
In the next section, we describe an implementation of an A2SC using a MAS approach and explore how these issues are handled.

\section{Prototype Development and Showcase}\label{sec:prototype}
The MIISI model, proposed in \citet{xu2022autonomous}, provides a conceptual framework for guiding A2SC design.
We illustrate A2SC implementation with a prototype model by following the MIISI framework and the MAS approach.

For a use case, we selected perishable food SCs, which tend to spoil because of improper storage and long transit times  during transportation.
Thus, it is important to monitor the ambient condition of the vehicle used for transporting perishable foods and rapidly respond to emergent events, such as traffic congestion. 

\begin{figure}[t]
    \centering
    \includegraphics[width=0.475\textwidth]{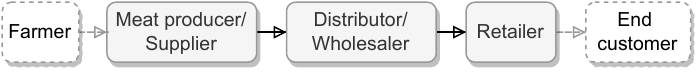}
    \caption{An illustration of a {\it simplified} meat supply chain.}
    \label{fig:meat_sc}
\end{figure}

The use case company (CMC) purchases meat from suppliers on a wholesale basis and supplies it to local restaurants. CMC would like to automate its wholesale and procurement procedures using  information and communications technologies such that the flow of information between different stakeholders can be accommodated. The prototype must maintain smooth movement of physical flow and its associated information flow concerning inbound and outbound goods shipment and state of inventory. The developed prototype  implements an {\it end-to-end} automated  procurement process using MAS, involving the set of stakeholders represented by agents along the supply chain. \figureautorefname\ref{fig:meat_sc} illustrates the main structure of the SC, in which our prototype focuses on the middle part of the chain (highlighted in grey). The prototype additionally includes two other stakeholders: logistics and third-party logistics (3PL), responsible for transporting foods and are essential to perishable foods SCs.

\subsection{Design}\label{sec:design}
The scenario has two primary processes: replenishment (CMC procures meat from its suppliers to replenish its inventory) and wholesale (retailer purchases meat from CMC). 
Both processes have overlapping functions, so we use the replenishment process for illustration.
Retailers can select their preferred delivery options from the multiple ones the logistics service offers retailers in the replenishment process.
To reduce redundancy, delivery in the wholesale process is assigned by logistics service providers.
Both processes also involve decision-making, such as making good proposals, proposal acceptance or refusal, and selecting a delivery option.
The prototype additionally contains functions to manage inventory, e.g., automatic stock updating and replenishment and logistic monitoring, where the real-time location and ambient conditions of products are monitored throughout the delivery.
In the next section,  we decompose the problem and distribute it to multiple agents.

\subsubsection{Agent decomposition and distribution}
Five stakeholders are involved in these two processes: the CMC, retailer, suppliers, logistics, and 3PL. An agent type represents each, as follows:
\begin{enumerate}[nosep]
    \item{Wholesaler agent}: 
    This agent type represents the wholesaler, CMC, managing the processes of procuring meat from supplier agents, wholesaling them to retailer agents, and managing its inventory.
    It  contacts logistics agents to arrange delivery services.
    It also serves as a central hub for this hypothetical SC, connecting upstream and downstream SC agents. 
    
    \item{Supplier agent}:
    This agent type manages the meat supply process to wholesaler agents. 
    It provides its customers with delivery services, which are in turn provided by its cooperated logistics agent. 
    The supplier agent thus coordinates with wholesale agents and logistics agents to complete a purchase order. 
    
    \item{Retailer agent}:
    This agent type represents local retailers, such as restaurants or local stores purchasing meat from wholesalers.
    It interacts with the wholesaler agent to deal with the procurement on behalf of a retailer. 
    
    \item{Logistics agent}:
    This agent type represents logistics service providers.
    It manages the overall operation of logistics services but outsources fulfilment services to 3PL providers. 
    It coordinates with suppliers, wholesalers, and 3PL providers for delivery arrangements.
    
    \item{3PL agent}:
    This agent, acting on behalf of a 3PL provider, is responsible for fulfilling delivery orders assigned by a logistics agent. 
    It monitors the entire transportation process of the meat, providing logistics companies with real-time data on the vehicle location and the product's ambient condition.
\end{enumerate}

All of the above agents are representative agents acting on behalf of SC stakeholders and are not involved in administrative tasks related to running the entire MAS.
Therefore, we need to add an{\ admin} agent to provide agent-related services, such as lookup and yellow page services.
These six types of agents are the main players whose interactions result in the dynamic structure of the prototype A2SC.

\subsubsection{Agent interaction, language, and protocols}
\begin{figure}[t]
    \centering
    \includegraphics[width=0.50\textwidth]{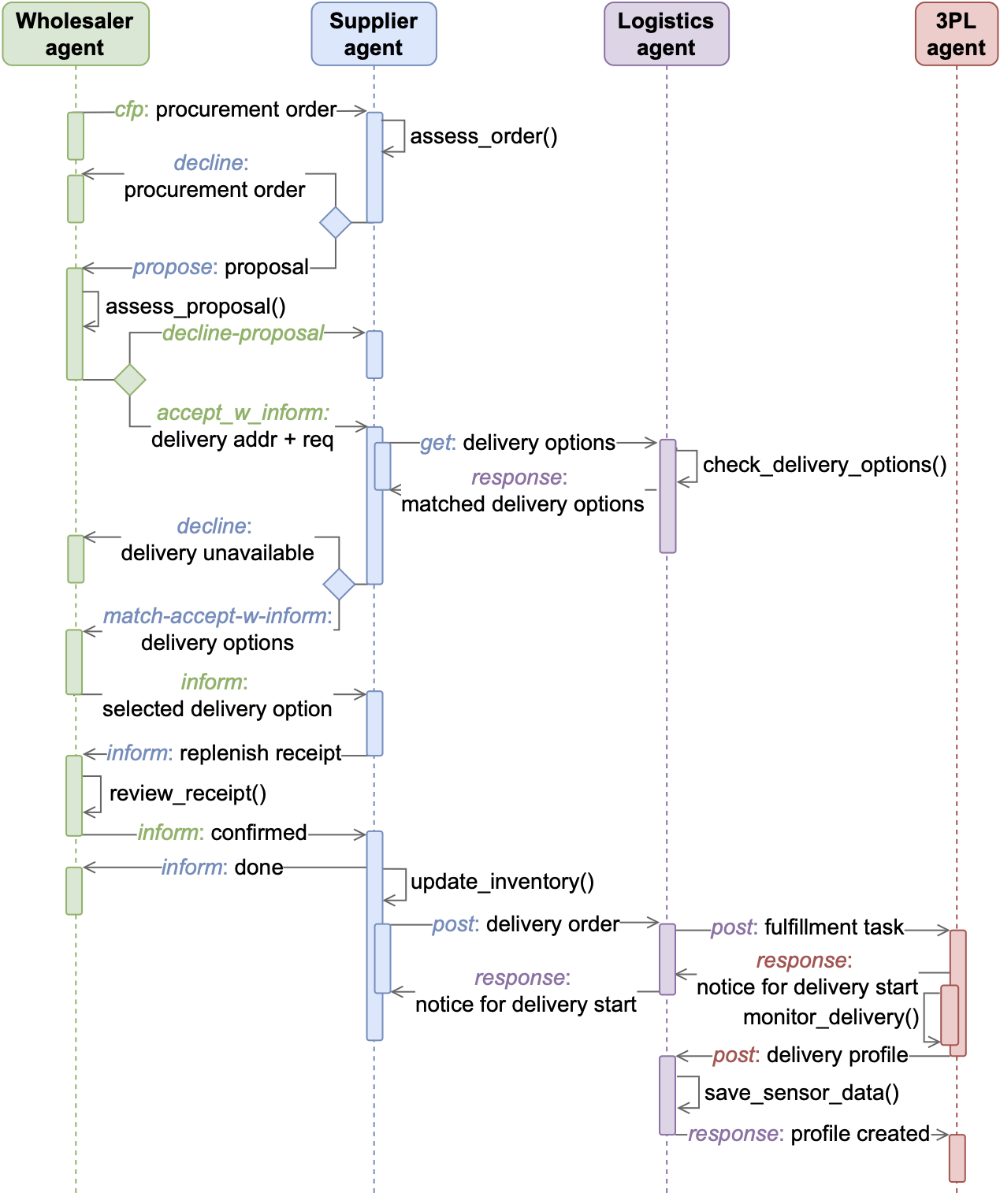}
    \caption{Agent interaction in the replenishment process.}
    \label{fig:agent_interaction_replenish}
\end{figure}
Agents communicate with other agents to complete certain tasks and must comply with commonly agreed interaction protocols to enable such interactions \citep{hosseini2021practical}. 
We consider two types of protocols in this prototype: contract-net and HTTP.
The contract net protocol is an interaction protocol for a service requester to seek suitable service providers who can perform certain tasks, which may involve multiple negotiation and information exchange rounds.
This protocol is suitable for handling complex interaction scenarios such as procurement which may involve unknown participants.
HTTP is a single-round interaction protocol adopted for handling simple interaction scenarios.
This protocol is suitable for interacting between known participants that have already established coherence, e.g., a logistics agent directly assigning a delivery task to one of its longstanding 3PL partners for fulfilment. As such, it is also appropriate for quickly requesting information, such as wholesale product prices, delivery rate cards, and current traffic conditions.

Interaction among agents was based on these two protocols.
For example, \figureautorefname\ref{fig:agent_interaction_replenish} illustrates agent interaction during the replenishment process, which involves four types of agents. As illustrated in (\figureautorefname\ref{fig:agent_interaction_replenish}), agents interact with each other through \emph{messages}. 
A message consists of two essential parts: the performative and the message body, delimited by a colon in \figureautorefname\ref{fig:agent_interaction_replenish}. 
These two parts denote the communicative act (behaviour) and the content of the message, respectively.

A set of performatives are defined in protocols for enabling interaction. 
For example, the contract-net protocol defines performatives such as \emph{cfp} (call for proposal), \emph{propose}, and \emph{reject-proposal}; 
the HTTP protocol has two performatives: \emph{request} (which in turn has two types of request: get and post) and \emph{response}.
For unambiguous communication, messages contain other information, such as sender, receiver, protocol, and ontology.
Agent communication languages define message format, and we use FIPA-ACL as the language in this study.

Both replenishment and wholesale processes involve multiple rounds of communication, sending and receiving messages to coordinate procurement and logistics services. 
They also involve procedures for supporting decision-making, such as \texttt{assess\_order()} for assessing incoming procurement orders, and self-calls for executing specific tasks, such as \texttt{monitor\_delivery()} for monitoring the delivery process.
Although the two processes are operationally separated and independent, their connection can be auto-triggered. 
For example, low wholesaler inventory stock triggers its replenishment behaviour and launches the replenishment process. 
In this sense, the two processes are linked, resulting in an end-to-end supply chain.

\subsection{Implementation}\label{sec:implementation}
Building on \sectionautorefname\ref{sec:design}, we adopted a client-server architecture to develop a web-based platform for this prototype.
This consists of three parts: backend MAS, frontend interfaces, and middleware.
We limit our discussion to the backend MAS as this defines the core model infrastructure.

\subsubsection{Open economic framework-based MAS}
\begin{figure}
    \centering
    \includegraphics[width=0.50\textwidth]{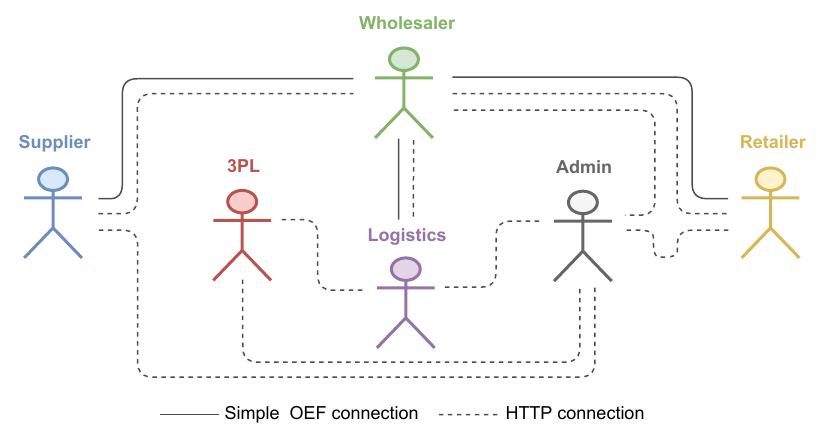}
    \caption{Illustration of the prototype's agent organisation.}
    \label{fig:agent_organisation}
\end{figure}

\begin{figure*}
    \centering
    \begin{subfigure}{0.48\textwidth}
        \includegraphics[width=\textwidth]{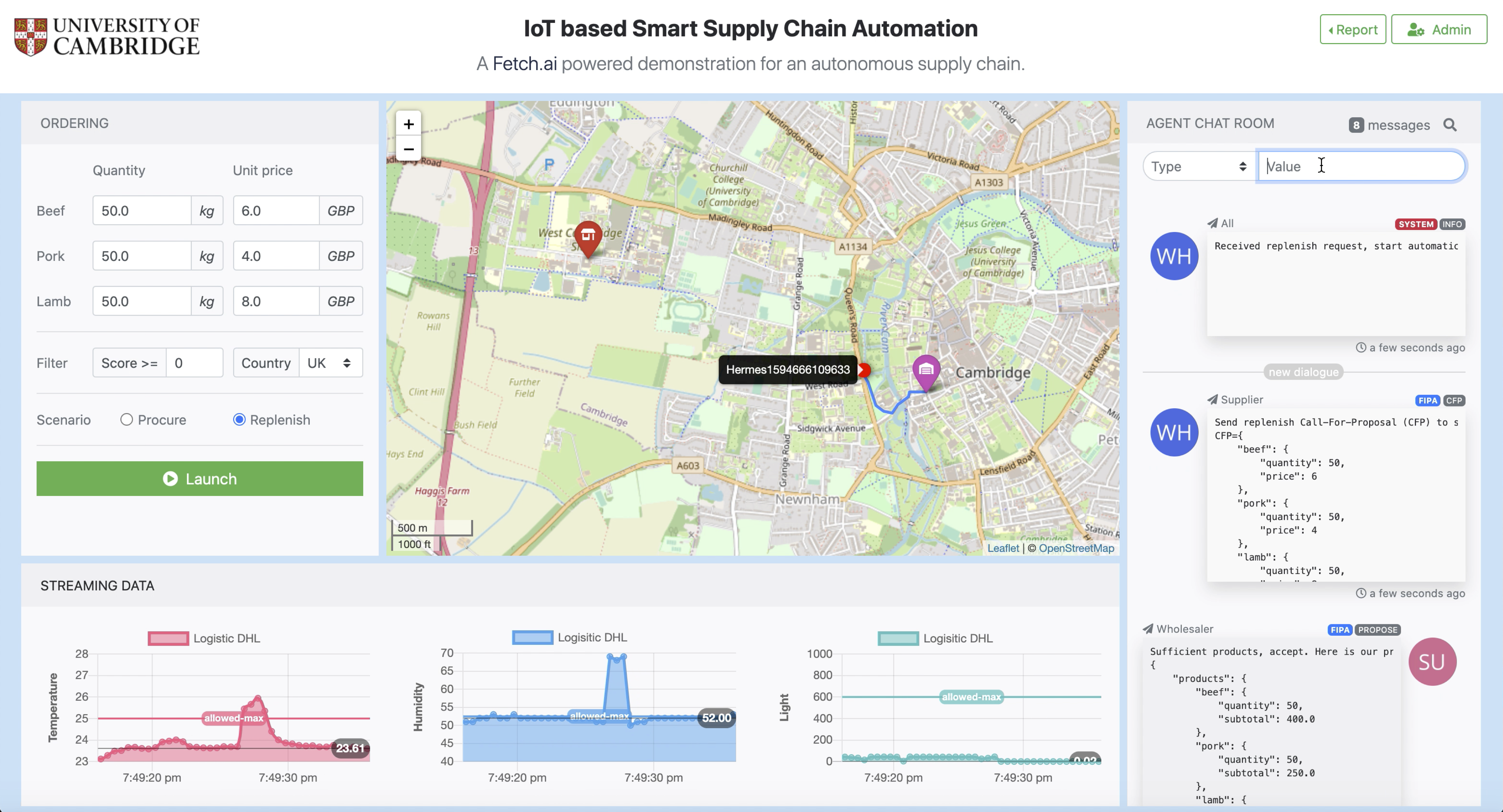}
        \caption{Replenish delivery in process.
              Relevant system notifications are shown at the top right of the interface; the supplier and wholesaler locations are marked on the map, being the delivery source and destination, respectively, with tracking number \texttt{Hermes1594666109633}.
              The dialogues among the agents are shown in the right panel.}
        \label{subfig:showcase_delivery}
    \end{subfigure}
    \hfill
    \begin{subfigure}{0.48\textwidth}
        \includegraphics[width=\textwidth]{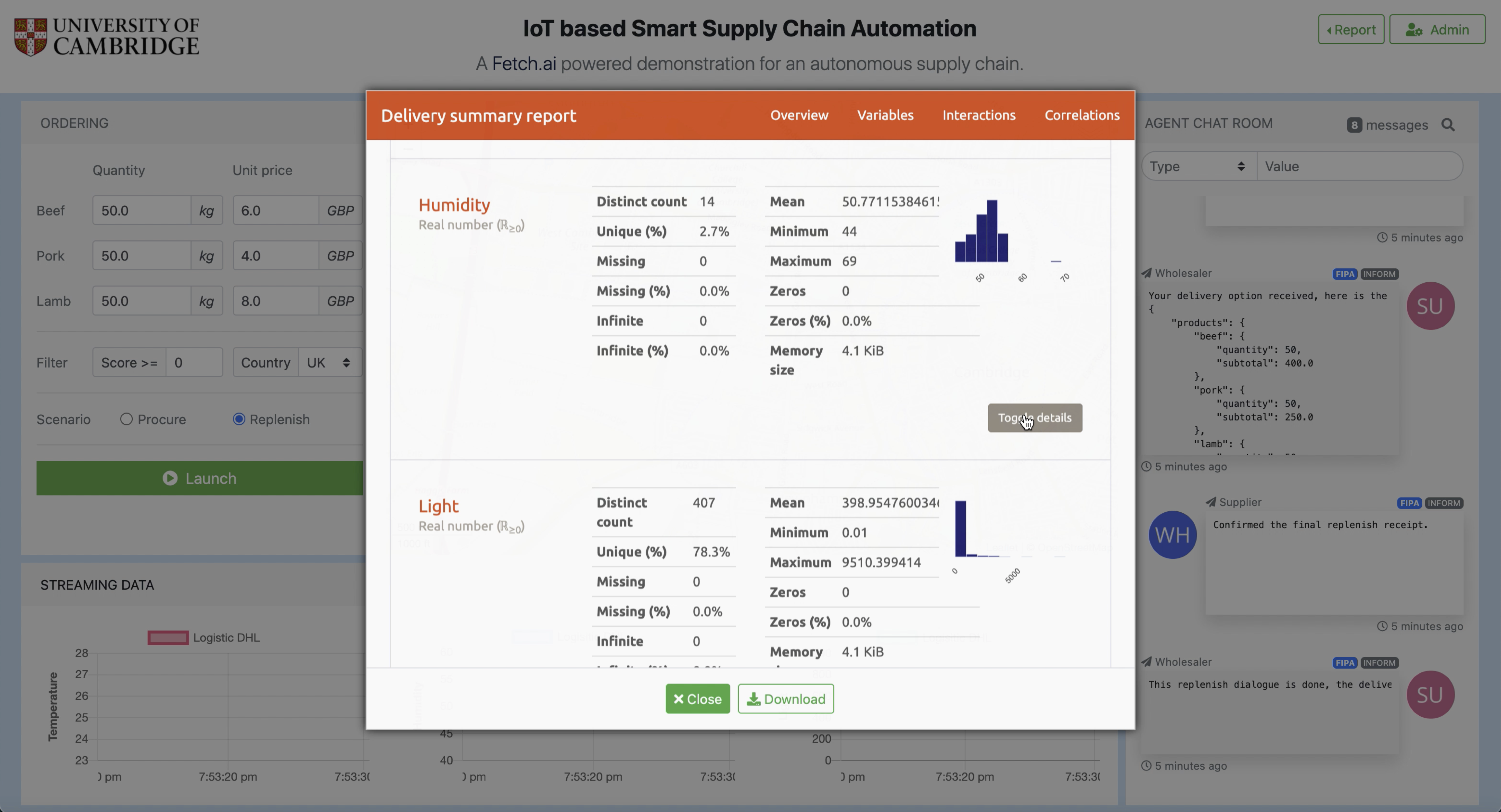}
        \caption{Replenish: delivery is done, summary report ready.
                 When the products are successfully delivered, a summary report based on the collected IoT data over the delivery journey can be accessed by clicking the `Report' button on the top right corner of the interface.
                 A summary report is shown in the centre scrollable pop-up window.}
        \label{subfig:showcase_summary}
    \end{subfigure}
    \caption{Example screenshots of the prototype.}
    \label{fig:screenshots} 
\end{figure*}
Agents in a MAS must be able to connect with each other, thereby achieving organised behaviour.
Thus a service requester must be able to find suitable service providers.
This is the \emph{connection} problem \citep{smith1981frameworks,ameri2013multi},
In this prototype, we realise agent connection through the open economic framework (OEF) \citep{minarsch2021autonomous}.
The OEF is a decentralised communication, search, and discovery system for their inhabitants --- autonomous economic agents (AEAs).
It consists of protocols, languages, and connection mechanisms, allowing agents to connect and then interact with each other.
The OEF hence provides a virtual environment --- a digital world --- for AEAs to live in by following social norms.

We used two types of connections supported by the OEF for the backend MAS: simple OEF and HTTP.
The simple OEF connection is an indirect (or mediated) connection, letting two unknown agents find each other via a mediator; 
whereas HTTP connection offers a direct connection between two familiar agents. 
Using an HTTP connection, agents can contact others via the request-response channel without the participation of the OEF.
The resulting agent organisation is shown in \figureautorefname\ref{fig:agent_organisation}.

\subsubsection{Implementation details}
This prototype continually collects real-time data to monitor the delivery process.
Due to the limited experimental conditions, it was not feasible to drive a vehicle with necessary data collection instruments and refrigerated containers to transport the perishable food product from one place to another and build wireless communication with remote agents.
We thus used the pre-collected IoT and GPS data to simulate real-time data collection during the delivery process.
To develop the prototype backend, we primarily used Python. 
We used other languages, such as HTML and JavaScript to create the frontend interfaces.
We also used the OEF, the AEA framework, Django, OpenStreetMap, etc. to facilitate the development of both the backend and frontend. 
We developed the agents using the AEA framework.

\subsection{Showcase}\label{sec:showcase}
After starting up the OEF, all of the agents, the Django built-in server, and the PostgreSQL service, the prototype is ready for use.
Each agent, after initialisation, first executes a registration behaviour. This one-shot proactive behaviour sends a message to the OEF to register their service to be discovered (connected) by other agents.
Users can access this prototype via a browser.
\figureautorefname\ref{subfig:showcase_delivery} and \figureautorefname\ref{subfig:showcase_summary} show the example screenshots of the prototype's interface.
The interface consists of four panels: 
ordering panel (top left), 
transport monitoring panel (central), 
agent chat room (right), 
and streaming data panel (bottom), as shown in \figureautorefname\ref{subfig:showcase_delivery}.

As the wholesaler's inventory is empty at the beginning of execution, we first replenish its inventory. 
We select the replenishment scenario and click the launch button with other default values. 
This  action triggers the wholesaler agent's replenishment behaviour.
The wholesaler agent first connects with the OEF, sending it a query message to search for suitable suppliers.
As a response, the recipient --- the OEF --- returns a message that contains a list of supplier agents that match the query conditions back to the sender.
This process is {\it matchmaking}, which matches a service requester with a service provider.
The wholesaler can then start a dialogue with the discovered suppliers and negotiate with a selected supplier on product purchase. 
The logistic agent also participates in this dialogue, in which the supplier requests available delivery options from this agent.
When these three parties agree on the deal, the logistics agent randomly selects a 3PL from a set of eligible 3PL agents to fulfil the delivery task.
The 3PL then begins the transport and monitors the whole journey. 
For example, \figureautorefname\ref{subfig:showcase_delivery} shows a prototype screenshot during the middle of this journey.
The relevant data about the ambient conditions are collected when the delivery is finished.
These data can be used for further analytics, for example, to determine product quality, as shown in \figureautorefname\ref{subfig:showcase_summary}.

\section{Discussion and Conclusions}\label{sec:discussion}
In this paper, we have explored the technical implementation of an autonomous supply chain system using a MAS approach after summarising MAS literature in supply chain management and highlighting the relationship between Digital Twinning and agent-based supply chain systems. 
We described four key components of an agent-driven autonomous supply chain system that need to be designed and  developed a concrete A2SC prototype to illustrate this approach, in which the Open Economic Framework (OEF) was adopted to facilitate supply chain agent connection and communication.
The OEF inhabitants --- i.e. agents --- were developed to act on behalf of SC entities. 
The approach illustrated the automation of a perishable goods procurement and wholesale process, consisting of ordering, coordination, transport monitoring, and follow-up analytics. 
As the automation of the processes themselves inherently requires real-time state information,  our prototype overlaps with extant definitions of Digital Twins in Supply Chains. Further research in Digital Twin and MAS integration would be beneficial in the context of supply chain information systems.
The prototype is a simple experimental proof of concept of an ASC based on the MIISI model \citep{xu2022autonomous} used to exemplify task decomposition, communication, and negotiation protocol implementation inherent in the design of an ASC.  
Its implementation focuses on the two middle layers of the model: \emph{interconnection} and \emph{integration}.
The prototype involves  {\it ten} agents, significantly less than what would be the actual case in a similar scenario setting. 
Thus future research needs to focus on the scalability of communications and resulting decision-making bottlenecks.
Moreover, the prototype's intelligence and automation are both low, being within the automation-skewed region of the autonomy manifold presented \citep{xu2022autonomous}.

Message flows enable interaction between agents within this structure, giving it cohesion. 
This approach also enables the integration of enterprise legacy systems, facilitating heterogeneous software such as ERPs to work together.  

As an initial attempt, this work seeks to shed light on implementing an ASC system in the real world, serving as a starting point for further studies.
Our future work includes involving more stakeholders, expanding the number of agents involved, and designing or introducing languages and protocols used explicitly for A2SCs.

\section*{Acknowledge}
This work was supported by the UK EPSRC Connected Everything Network Plus under Grant EP\slash S036113\slash 1 and the DO-TES project.

\bibliography{references}

\end{document}